**Swift detects a remarkable gamma-ray burst, GRB 060614, that introduces a new classification scheme**

N. Gehrels[1], J.P. Norris[1], V. Mangano[2], S.D. Barthelmy[1], D.N. Burrows[3], J. Granot[4], Y. Kaneko[5], C. Kouveliotou[6], C.B. Markwardt[1,7], P. Mészáros[3,8], E. Nakar[9], J.A. Nousek[3], P.T. O'Brien[10], M. Page[11], D.M. Palmer[12], A.M. Parsons[1], P.W.A. Roming[3], T. Sakamoto[1,13], C.L. Sarazin[14], P. Schady[11,3], M. Stamatikos[1,13], & S.E. Woosley[15]

[1]*NASA/Goddard Space Flight Center, Greenbelt, MD 20771, USA*

[2]*INAF -- Osservatorio Astronomico di Brera, Via Bianchi 46, 23807 Merate, Italy*

[3]*Department of Astronomy & Astrophysics, Pennsylvania State University, PA 16802, USA*

[4]*KIPAC, Stanford University, P.O. Box 20450, MS 29, Stanford, CA 94309, USA*

[5]*USRA, NSSTC, VP-62, 320 Sparkman Drive, Huntsville, Alabama 35805, USA*

[6]*NASA/Marshall Space Flight Center, NSSTC, VP-62, 320 Sparkman Drive, Huntsville, Alabama 35805, USA*

[7]*Department of Astronomy, University of Maryland, College Park, MD 20742, USA*

[8]Department of Physics, Pennsylvania State University, PA 16802, USA

[9]Theoretical Astrophysics, California Institute of Technology, MS 130-33, Pasadena, CA 91125, USA

[10]Department of Physics and Astronomy, University of Leicester, Leicester LE1 7RH, UK

[11]Mullard Space Science Laboratory, University College London, Dorking, RH5 6NT, UK

[12]Los Alamos National Laboratory, Los Alamos, NM 87545, USA

[13]Oak Ridge Associated Universities, P.O. Box 117, Oak Ridge Tennessee 37831-0117, USA

[14]Department of Astronomy, University of Virginia, Charlottesville, VA 22904-4325, USA

[15]Department of Astronomy and Astrophysics, University of California at Santa Cruz, Santa Cruz, CA 95064, USA

**Gamma ray bursts (GRBs) are known to come in two duration classes[1], separated at ~2 s. Long bursts originate from star forming regions in galaxies[2], have accompanying supernovae (SNe) when near enough to observe and are likely caused by massive-star collapsars[3]. Recent observations[4-10] show that short bursts originate in regions within their host galaxies with lower star formation rates,**

**consistent with binary neutron star (NS) or NS - black hole (BH) mergers[11,12]. Moreover, although their hosts are predominantly nearby galaxies, no SNe have been so far associated with short GRBs. We report here on the bright, nearby GRB 060614 that does not fit in either class. Its ~102 s duration groups it with long GRBs, while its temporal lag and peak luminosity fall entirely within the short GRB subclass. Moreover, very deep optical observations exclude an accompanying supernova[13-15], similar to short GRBs. This combination of a long duration event without accompanying SN poses a challenge to both a collapsar and merging NS interpretation and opens the door on a new GRB classification scheme that straddles both long and short bursts.**

The Burst Alert Telescope (BAT) onboard the Swift satellite[16] detected GRB 060614 on 2006 June 14 at 12:43:48 UT. Based on the initial positions from the on-board X-ray and UV/optical telescopes (XRT, UVOT), the burst was subsequently located by ground- and space-based telescopes at the outskirts of a relatively nearby faint dwarf galaxy at a redshift of $z=0.125$[17]. We find the suggestion[18,19] of a chance alignment between a background GRB and foreground galaxy at z=0.125 to not be credible; the chance probability of the observed 0.5" offset between the GRB and the z=0.125 galaxy to be by chance is only $2 \times 10^{-5}$. Also, fits to the combined UVOT and XRT spectra give $z<1.3$ at the 99.99% confidence level, excluding the suggested[18] location at z>1.4. (See Gal-Yam et al.[14] for additional evidence against a chance alignment.)

The event duration, $T_{90}$ = 102 s (15-350 keV; where $T_{90}$ is the time during which 90% of the event photons were collected), clearly places GRB 060614 in the long burst category[1]. As the event was also at close proximity, it became a prime candidate for a SN search in its light curve, as indeed had been found in four individual cases in the past[20]. The three accompanying papers[13-15] report very tight upper limits -- 100 times lower than previous detections -- in these searches. Thus far, *strict* SN limits had been found only for the GRBs of the short variety. The question then is: why was there no supernova associated with the long GRB 060614?

Interestingly, a close examination of the BAT light curve of GRB 060614 (Figure 1) reveals a first short, hard-spectrum episode of emission (lasting 5 s) followed by an extended and somewhat softer episode (lasting ~100 s). The total energy content of the second episode is five times that of the first [fluence of $(1.69\pm0.02)\times10^{-5}$ erg cm$^{-2}$ and $(3.3\pm0.1)\times10^{-6}$ erg cm$^{-2}$, respectively, in the 15-350 keV band]. The recent discovery of a similar two-component emission structure in several *Swift* and *HETE-2* short GRBs prompted Norris and Bonnell[21] to search the BATSE GRB database for similar events. They found 3 bursts with a *bright* softer emission component, spectrally similar to that of GRB 060614 (corresponding to roughly ~1% of the number of short BATSE GRBs). Weaker soft components are found in 4 out of 16 *Swift* and *HETE-2* short bursts and 11 out of 130 *Konus* short GRBs[22], a significant ~10%-25% of their short GRB databases. In fact, *it may be that the majority of short bursts have such soft components*; currently, only 2 *Swift* short GRBs (051221A, 060313) were bright enough to allow low fractional limits to be set. It is interesting to note that regardless of their duration, almost all of these events with softer components have a common appearance, namely a 5 - 10 s gap between the first and second episodes and a humped shape of the second component.

Another method to distinguish categories of GRBs is to compute the temporal lags between their light curve features in different energy bands[21,23,24]. We introduce a new variant on this approach by including short bursts in a plot of peak luminosity ($L_{peak}$) versus lag ($t_{lag}$); the redshift measurements from the past year for a number of short *Swift* GRBs makes the comparison possible for the first time. There is an anti-correlation between lag ($t_{lag}$) and peak luminosity ($L_{peak}$) for long bursts as shown in Figure 2. In contrast, short bursts have small $t_{lag}$ and small $L_{peak}$ and occupy a separate area of parameter space from long bursts. The lag for GRB 060614 for the first 5 s is 3 ± 6 ms. Thus, in spite of its long duration, GRB 060614 falls in the same region of the lag-luminosity plot as short bursts. Moreover, we were able for the first time to accurately calculate a lag for the softer second episode due to its spiky light curve. We find $t_{lag}$=3±9 ms, consistent with the short-hard episode, and thus indicating a similar origin. The physical cause of lags in GRBs is not yet well understood, but it has been suggested[23] that bursts with smaller lags have more relativistic outflows.

From the above, it is difficult to determine unambiguously which category GRB060614 falls into. It is a long event by the traditional definition, but it lacks an associated SN as observed in other nearby long GRBs, and it shows morphological similarities with *Swift* short bursts. Further, the complex structure of the second component and similar lag with the first component indicates that *the extended episode is due to the long-lived activity of the burst central engine rather than due to the onset of the afterglow*

*emission.* However, its spectral lag and low luminosity make it distinct from other long GRBs and place it together with short burts. Perhaps the odd characteristics of GRB 060614 point to a new subclass with a different physical origin identified by a third GRB property, namely spectral lag. However, we first must ask what it would take to fit such a class of events into either the collapsar or merger models.

Interestingly, short-hard bursts with continuing activity resembling long-soft ones are a prediction of some massive star models[25], favoured when the pre-explosive mass loss is unusually high and the jet energy low. The lack of a SN requires that significant production of $^{56}$Ni is avoided during the star collapse. This might be possible if the disk wind is weak and if $^{56}$Ni that had been produced initially fell back in a weak supernova explosion. Nevertheless, producing less than $10^{-3}$ $M_O$ of $^{56}$Ni is a challenge for the collapsar model[26]. Another alternative is a failed SN; the small observed lag may imply a higher velocity outflow with insufficient energy deposited in the star for it to explode. In the NS merger model the duration is set by the lifetime of the accretion disk around the newly born BH and is expected[27] to be ~0.1 s. The accretion during a BH-NS merger might be longer if, as suggested by simulations[28], a significant amount of mass is ejected to a bound orbit and falls back on the BH on time scales larger than 1 s. In short, GRB 060614 may be either just another odd event in the motley-crew collection of GRBs, or, more likely, one of the first established members of an emerging new class of events that include both short and/or long GRBs.

Reprints and permission information is available at npg.nature.com/reprintsandpermission. The authors declare no competing financial interests. Correspondence and requests for materials should be addressed to N. Gehrels (gehrels@milkyway.gsfc.nasa.gov).

Figure 1 **The light curve of GRB 060614 as observed with the BAT**. The mask-tagged BAT light curve is shown in 4 energy bands in the top panels, with a sum in the bottom panel. The outtake panel gives an expanded view of the first episode. There is a hint of a 9 s periodicity between 7 and 50 s, but we find it to be not statistically significant. The spectrum is well fitted by a power-law in both intervals with a photon index of 1.63±0.07 in the first 5 s and a softer inndex of 2.02±0.04 in the following 120 s. The burst was also detected with Konus-WIND[29] with $E_{peak}$ = 302 (-83, +214) keV in the first 5 s. The total energy radiated assuming isotropic emission, $E_{iso}$, in the 1 keV - 10 MeV range in the GRB restframe for the first 5 s is $1.8 \times 10^{50}$ erg. We note that the $E_{peak}$-$E_{iso}$ value for the short episode falls far off of the long-burst Amati relationship. The spacecraft repointed XRT and UVOT at the burst in 90 s. The X-ray emission (0.3-10 keV) was initially among the brightest to date at $5 \times 10^{-8}$ erg cm$^{-2}$ s$^{-1}$ at 100 s. The afterglow light curves suggest a jet break at ~1.5 days, implying a narrow jet occupying $\leq 10^{-2}$ of the total solid angle and suggesting an energy of ~$10^{49}$ erg in the relativistic jet, roughly comparable to that inferred for *Swift* short GRBs.

Figure 2  **Spectral lag as a function of peak luminosity showing GRB 060614 in the region of short GRBs.**  The lags and peak luminosities are corrected to the source frame of the GRB.  The lags are defined as the time difference between light curve structures in the 50-100 keV and 15-25 keV channels.  The energy correction (K-correction) to the source frame for the lag is approximately $(1+z)^{0.33}$, making the total energy correction plus time dilation correction $(1+z)^{-0.67}$.  The data points labelled as long bursts are from *Swift/BAT*, with the exception of GRB 030528 which is a very long-lagged *HETE-2* burst.  The dashed line illustrates the lag-luminosity correlation for long bursts.  Outliers are bursts with unusual properties: GRB 060729 with extremely long-lived afterglow and 051016B with extremely soft prompt emission.  The blue data points for short bursts are from *Swift/BAT*.  The peak luminosity was calculated for the short bursts and GRB 060614 using 64 ms binning.  In green are the 4 nearby long GRBs with associated SNe.  The three of the four SNe-associated underluminous nearby GRBs (980425, 031203 and 060218) fall below the long-burst correlation, while the only SN-associated GRB with normal luminosity (030329) falls near the line giving further evidence of it being a normal GRB (the only normal GRB with a firm SN association).  None of the SN-associated GRBs falls near the short grouping.  Note that our detailed calculation of the lag of GRB 030329 gives a larger value here than previously found[30] in a higher energy band.

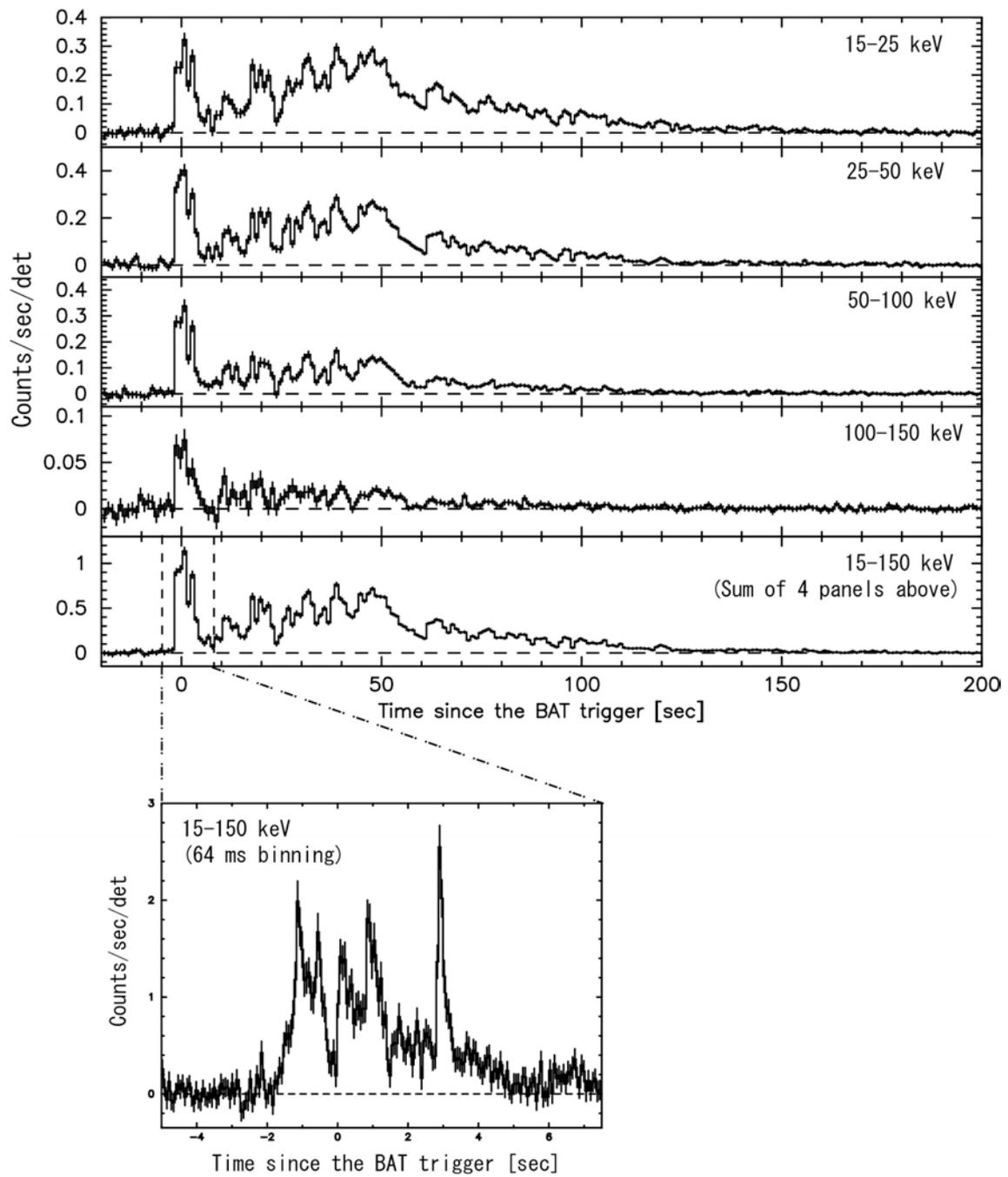

**Figure 1**

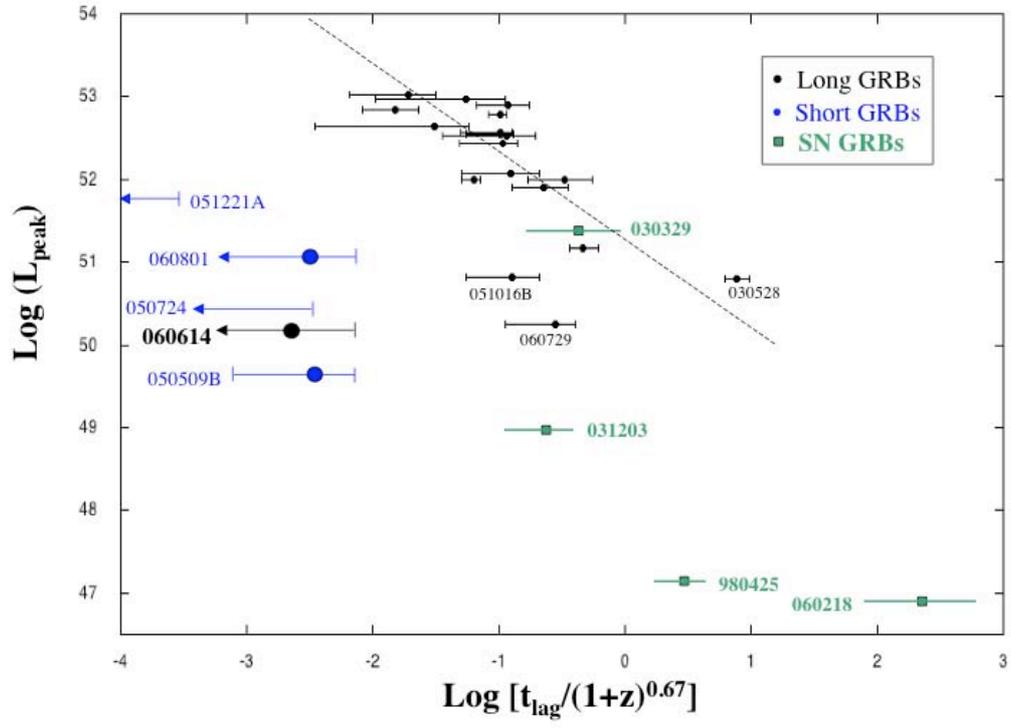

**Figure 2**